\documentclass{article}
\usepackage[utf8]{inputenc}

\usepackage{graphicx}  
\usepackage{color}
\usepackage{color,soul} 
\usepackage{dcolumn}   
\usepackage{bm}        
\usepackage{longtable}
\usepackage{amssymb,amsmath}   
\usepackage{fullpage,cite} 
\usepackage{hyperref}
\usepackage{times}
\usepackage[T1]{fontenc}
\usepackage{amsthm,amssymb,amsfonts,amsmath,graphicx,cancel,multirow,array,float}
\usepackage{colortbl,feynmf}

\title{Some remarks on projectile motion with a linear resistance force}
\date{}

\author{C. A. Morales$^{1,2}$, J. H. Muñoz$^{1, \dagger}$ , C. E. Vera$^{1}$}

\begin{document}

\maketitle

\begin{center}
$^{1}$Departamento de F\'isica, Universidad del Tolima, C\'{o}digo Postal 730006299, Ibagu\'{e}, Colombia \medskip
$^{2}$ Institución Educativa Francisco de Miranda, Rovira-Tolima, Colombia  \medskip
$^{\dagger}$ Corresponding author
\end{center}

\begin{abstract}

In this article we revisit the projectile motion assuming a retarding force proportional to the velocity, $\vec{F_r} = -mk\vec{V}$.  We obtain an analytical expression for the set of maxima of the trajectories, in Cartesian coordinates, without using the Lambert $W$ function. Also, we investigate the effect of the parameter $k$ on  the radial distance of the projectile showing that  the radial distance oscillates from a certain critical launch angle  and find an approximate expression for it. In our analysis, we consider the impact of the parameter $k$ in the kinetic energy, the potential energy, the total energy and the rate of energy loss, and in the phase space. Our results  can be included in an intermediate-level  classical mechanics course.  \medskip


\end{abstract}

\section{Introduction}

The projectile motion in a constant gravitational field is an important topic that is studied  in  introductory physics  courses at university level. It is considered in absence of medium resistance in  almost all  fundamental physics  textbooks (see for example \cite{Serway2004, Halliday2010, Tipler2008, Sears2008}). To consider a more realistic situation, it is necessary to include    retarding forces ($F_r$).   A good approximation, in this case,  is to assume that they are  proportional to some power of the speed ($F_r \propto v^n$) \cite{Appell, Marion-Thornton, Symon, Murphy1972, Mestre1990, Lange, Richard}.   \medskip

In this paper,  we revisit the motion of a projectile in a fluid  considering the effect of a   retarding force proportional to the velocity ($n=1$): $\vec{F_r} = -mk\vec{V}= -mk(\dot{x}  \, \widehat{i}+\dot{y}\, \widehat{j} )$, where  $\dot{x} = dx/dt$, $\dot{y} = dy/dt$, $m$ is the mass of the projectile and $k$ is a positive constant that specifies the strength of the resisting force. The unit of $k$ is \textcolor{blue}{s}$^{-1}.$ The linear drag model, $\vec{F_r} \sim \vec{V}$, is a good approximation when the dimensionless Reynolds number is small, indicating that the inertia forces are negligible with respect to the viscous forces and the fluid  has a  laminar flow. Under these conditions,  the Stokes' law is valid for a sphere of radio $R$ moving in a fluid and $k = 6 \pi R \eta$, where  $\eta$ is the viscosity of the medium  \cite{Vennard, Bachelor, Long-Weiss}.  \medskip

The projectile motion with linear resistance force has been extensively discussed in the literature a long time ago  using different approaches. According to our bibliographical review, this literature can be classified in two groups. In one of the them, the studies were performed by means of approximate methods, whereas in the other group  the researches  were performed by introducing the Lambert $W$ function in order to obtain analytical expressions. \medskip

In References \cite{Marion-Thornton, Symon, Murphy1972, Mestre1990, Greiner2003, Erlichson1983, Martin1991, Groetsch1996, Groetsch1997, Alwis2000, Bruno2002,  Groetsch2003, Groetsch2005, Fowles2005, Pereira2008, Borghi2013, Andersen2015, Kantrowitz2015, Grigore2017, Rizcallah2018, Pispinis2019, Sarafian2021, Ribeiro2021, Hernandez2022} it is obtained, by means of approximate procedures or through  computational tools,  the trajectory of the particle, the time of flight, the maximum height,  the range, the curve of safety or the path length. On the other hand, in the last two decades the interest in the projectile motion with a retarding force proportional to the   velocity has increased because   it is a good scenario to apply the Lambert \textit{W} function since it is necessary to solve transcendental equations in this problem. In this direction,  several authors have obtained analytical expressions for the shape of the trajectory,  the range, the angle that gives  the maximum range (the optimal launch angle), the time of flight, the time of ascent (or descent),  the time of fall, the maximum height, the locus of the apexes (in Cartesian  and polar coordinates) in terms of  the Lambert $W$ function \cite{Warburton2004, Packel2004, Stewart2005, Stewart2005A, Morales2005,Stewart2006, Kantrowitz2008, Karkantzakos2009, Hernandez2010, Stewart2011Comment, Hernandez2011Reply, Stewart2011, Morales2011, Stewart2012, Hu2012, Bernardo2015, Morales2016}.   \medskip

Stewart \cite{Stewart2006} and Hernandez-Saldaña \cite{Hernandez2010} obtained, using the Lambert $W$ function,  the locus of the set of apexes corresponding to the  maximum heights in Cartesian and polar coordinates, respectively, for the projectile motion with a retarding force proportional to the   velocity. Motivated by these two works, we have extended them obtaining an analytical expression for this locus in Cartesian coordinates without introducing the Lambert $W$ function, through  a simple and didactical procedure, with the help of  \textit{Mathematica} (as in Ref. \cite{Alwis2000}). According to our knowledge, this result has not been previously reported. \medskip

 In addition, we have scrutinized the effect of the parameter $k$ on  the radial distance of the projectile motion with a retarding force proportional to the   velocity, motivated by the previous works of  Walker \cite{Walker1995} (without friction) and, Ribeiro and Sousa  \cite{Ribeiro2021}. Our results complement these two works. Moreover, we have performed a pedagogical and didactical overview on typical observables associated to the projectile motion with a friction force proportional to the   velocity, including in our analysis observables as kinetic energy, potential energy, total energy an the rate of energy loss that  have not been considered in the literature up to now.  \medskip

The paper is organized as follows. In section 2 we present an overview on several observables of the projectile motion with a retarding force proportional to the velocity,  including the kinetic energy, the potential energy, the total energy,  the rate of energy loss and the trajectory in the phase space. In section 3,  we         analyze the impact of the parameter $k$ on the radial distance and the critical angle for obtaining a radial oscillation. In section 4, the most important,  we study the evolution  of the apexes of the trajectories in function of the launch angle, in Cartesian coordinates without using the Lambert $W$ function. Finally, in section 5 we summarize our principal results. 

\section{General results: An overview}

Let us assume that in $t=0$ the projectile is launched from the origin of the coordinate system with the initial velocity $\vec{V_{0}}$ and the angle of elevation $\theta_0$. The equations of motion in the horizontal and vertical directions, respectively, are
\begin{eqnarray}
m \ddot{x} &=& -  m k \dot{x},   \\
m \ddot{y} &=& - m k\dot{y} - mg,
\end{eqnarray}
where $m$ is the mass of the projectile,  $g$ is the acceleration of  gravity,  $\dot{x}$ ($\dot{y}$) is the horizontal (vertical) velocity  and  $\ddot{x}$ ($\ddot{y}$) is the horizontal (vertical) acceleration of the projectile. These differential equations are solving making $\ddot{x} = d\dot{x}/dt$ and  $\ddot{y} = d\dot{y}/dt$ and integrating to obtain $\dot{x}$ and $\dot{y}$ in function of time. The solutions for these equations are well known  \cite{Marion-Thornton, Murphy1972, Mestre1990,  Erlichson1983, Martin1991, Groetsch1996, Groetsch1997, Alwis2000, Bruno2002, Groetsch2005, Fowles2005, Pereira2008, Grigore2017}:

\begin{equation}\label{position-x}
x(t) = \dfrac{U}{k} \left( 1 - e^{-kt} \right), 
\end{equation}

\begin{equation}\label{position-y}
y(t) = -\dfrac{gt}{k} + \dfrac{kV + g}{k^{2}}\left( 1 - e^{-kt}       \right),
\end{equation}

where $U = V_{0} \cos \theta_0$ and $V = V_{0} \sin \theta_0$.  

The horizontal and  vertical components of the velocity are
\begin{equation}\label{velocity-x}
\dot{x}= U e^{-kt}, 
\end{equation}

\begin{equation}\label{velocity-y}
\dot{y} = -\dfrac{g}{k} + \dfrac{kV + g}{k}e^{-kt}, 
\end{equation}

and the components of the acceleration are  given by
\begin{eqnarray}\label{acceleration}
\ddot{x} &=& -kU e^{-kt},    \\
\ddot{y} &=& -(kV + g)e^{-kt}.
\label{acceleration1}
\end{eqnarray}

We will now investigate the phase space of projectile motion under the influence of a linear resistance force, which has not been studied so far. Our aim is to enhance our understanding of the temporal evolution of some variables and find recurring patterns.\medskip

\begin{figure*}[!t]
	\centering
 \includegraphics[scale=0.375]{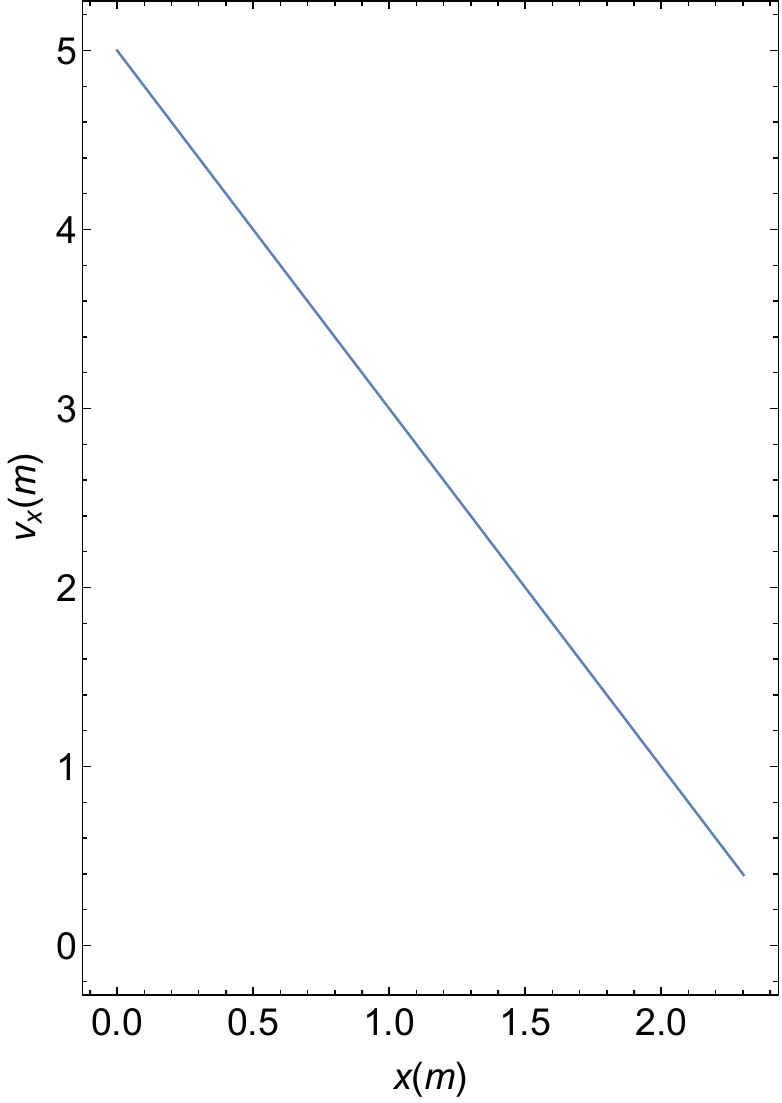} \hspace{0.36cm}
  \includegraphics[scale=0.39]{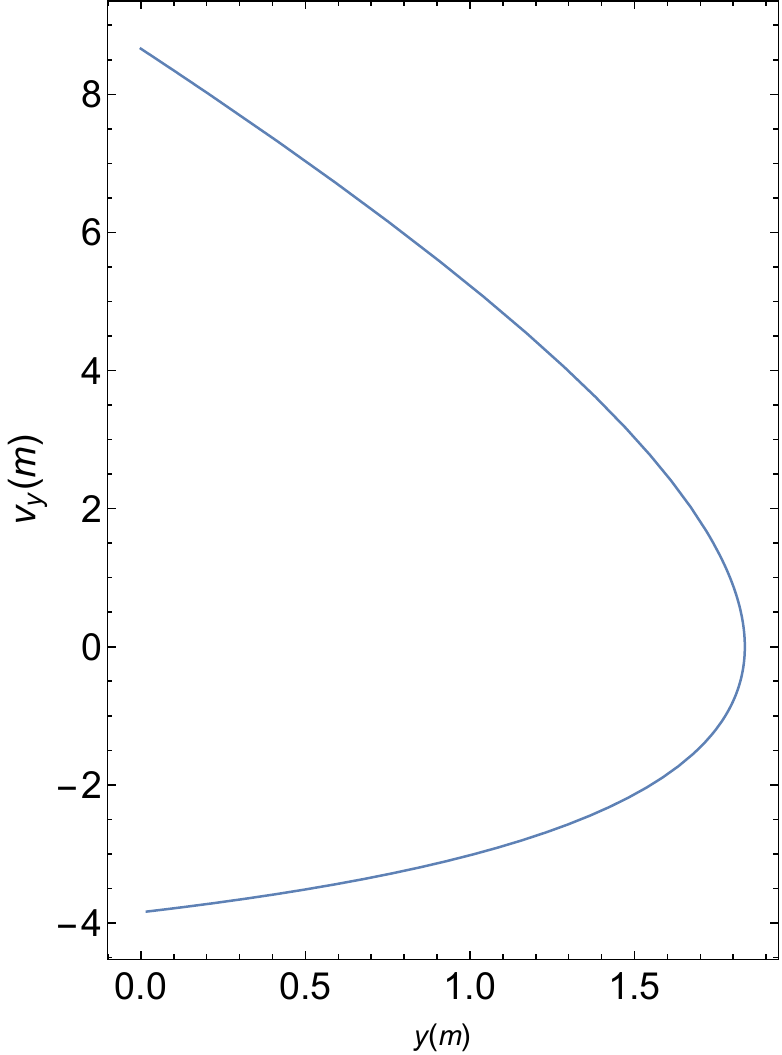}\hspace{0.36cm}
 \includegraphics[scale=0.384]{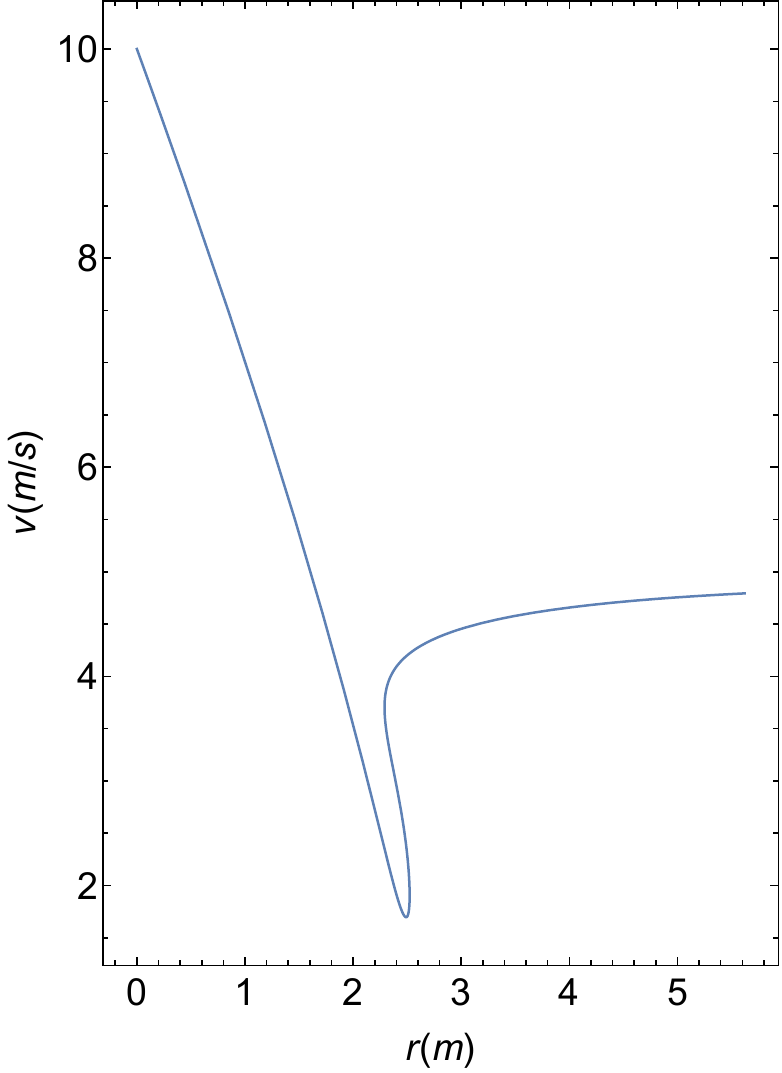}\vspace{-0.15cm} 
	\caption{\small Phase plots $\dot{x}$ vs $x$ \textit{(left)}, $\dot{y}$ vs $y$ \textit{(middle)} and $v$ vs $r$ \textit{(right)} with $k = 2$  s$^{-1}$, $V_0 = 10$ m/s and $\theta_0 = 60^{\circ}$.}
	\label{phase space}
\end{figure*}

The trajectory  in the phase space $\dot{x}$ vs $x$ is obtained from expressions \eqref{position-x} and \eqref{velocity-x} for $x(t)$ and $\dot{x}(t)$, respectively. It is:  
 \begin{equation}\label{x-Vx}
     \dot{x} = U - kx.
 \end{equation}
This trajectory corresponds to a straight line (Figure \ref{phase space} \textit{(left)}). Its slope is the resistance coefficient $k$. Experimentally, this expression could be useful for obtaining the parameter $k$ of a laminar fluid by plotting $\dot{x}$ versus $x$.  \medskip

In a similar way, the trajectory  in the phase space $\dot{y}$ vs $y$ is obtained from  equations \eqref{position-y} and \eqref{velocity-y} for $y(t)$ and $\dot{y}(t)$, respectively,  giving
\begin{equation}\label{y-Vy}
   \frac{g}{k} \text{ln} \left[ \frac{k}{kV + g}\left(\dot{y} + \frac{g}{k}  \right)   \right] - \dot{y} = ky - V.
\end{equation} 

The Figure \ref{phase space} \textit{(middle)} shows this phase plot.  When there is no friction, it is a parabola.   We can see that the graph is not symmetric with respect to the line $\dot{y}=0$ because during the motion, some of the projectile's mechanical energy dissipates, converting into thermal energy as a result of collisions between the molecules composing the fluid and the projectile. Therefore, for a given height, there are two values of $\dot{y}$ that correspond to when the projectile goes up and goes down, such that $|\dot{y}(going \ up)| > |\dot{y}(going \ down)|.$  \medskip

Additionally, in the Figure \ref{phase space} \textit{(right)} it is displayed the phase plot $v$ vs $r$, where $v = \sqrt{\dot{x}^2 + \dot{y}^2}$ is the magnitude of the velocity and $r = \sqrt{x^2 + y^2}$ is the radial distance.   The minimum value of $v$ does not correspond to the point of maximum height. From the radial distance at which $v$ is minimum, the projectile falls, increasing its vertical velocity and decreasing its horizontal velocity until it reaches a constant value equal to $g/k$. As a result, the curve asymptotically tends towards the line $v=g/k$. For $k=2$, this value is approximately $(9.8 \ \text{m/s}^2)/(2 \text{s}^{-1}) = 4.9 \ \text{m/s}$ (This value can be verified in  Figure \ref{phase space} \textit{(right)}). \medskip

Now we are going to get the time required for the projectile to arrive at the  maximum height ($= t_{peak}$). Taking $\dot{y} = 0$  in equation \eqref{velocity-y} it is  obtained that the projectile reaches the peak at the time
\begin{equation}\label{t-subida}
t_{peak} = \dfrac{1}{k} \text{ln} \left[  \dfrac{kV + g}{g}     \right].
\end{equation}

The maximum height $(Y_{m})$ and the horizontal position corresponding to this height 
$(X_{m})$ are obtained substituting equation \eqref{t-subida} in $y(t)$ and $x(t)$, respectively:
\begin{equation}\label{Ymax}
Y_{m} =  -\dfrac{g}{k^{2}} \text{ln} \left[ 1 +  \dfrac{kV}{g} \right] + \dfrac{V}{k}  
\end{equation}

\begin{equation}\label{Xmax}
X_{m} =   \dfrac{UV}{g}\left( 1 + \dfrac{kV}{g} \right)^{-1}.  
\end{equation}

The point $(X_m, Y_m)$ gives the apex of the trajectory. We will discuss in section 4 the evolution of these points as a function of the launch angle.  \medskip

Now let us get the trajectory of the projectile. To do this, we obtain the time $t$ from the equation $x=x(t)$:
\begin{equation}
t = \dfrac{1}{k} \text{ln} 
\left[ \left(   1 - \dfrac{k}{U}x \right)^{-1}\right], 
\end{equation}
and substitute it  in $y = y(t)$:
\begin{equation}\label{Eq-trajectory}
y =  \dfrac{g}{k^{2}}  \text{ln}\left(   1 - \dfrac{k}{U}x \right) + \dfrac{kV+g}{kU}x.
\end{equation}

The range $R$ of the projectile is obtained by setting  $y = 0$ in equation \eqref{Eq-trajectory}. By doing this, it is obtained

\begin{equation}
\frac{g}{k^2}  \text{ln} \left( 1 - \frac{k}{U}R \right) + \frac{V + \frac{g}{k}}{U}R = 0.
\end{equation}

Obtaining an analytical solution for the range $R$ from the last equation is not easy because in this expression a linear function for $R$ is equated to a logarithmic function for $R$. However, it is possible to obtain an approximate expression for it using the expansion $\text{ln}(1+z) = z - z^{2}/2 + z^{3}/3 - ...$. By doing this, it is obtained:
\begin{equation}
R \approx \dfrac{2UV}{g}  \left( 1 - \dfrac{kV}{4g} \right).   
\end{equation}

For $k = 0$, we obtain $R \approx 2UV/g = V_0^2 \sin 2\theta_0 /g$, which is the well-known expression for the case where there is no friction. \medskip

Next, let us consider the  time of flight $(T_f)$, which is the  required time  for the projectile   returns to the ground. It is obtained taking $y=0$ in $y(t)$ (see equation \eqref{position-y}. Doing this, it is obtained 
\begin{equation}\label{timeflight1}
gT_f - \dfrac{kV + g}{k}\left(  1 - e^{-kT_f}       \right)=0.
\end{equation}
Using the expansion $e^z = 1 + x + x^2/2 + \ldots$ and considering only terms through $k^2$ in the last equation, the time of flight  is, approximately,  
\begin{equation*}
 T_f \approx \frac{2V}{kV + g} = \frac{2V}{g(1 + kV/g)}= \frac{2V}{g}(1 + kV/g)^{-1},     
\end{equation*}

\begin{equation}\label{timeflight2}
 T_f \approx  \frac{2V}{g}(1 - kV/g).     
\end{equation}
  This result is  a good approximation only for  small values of $k$. If $k = 0$, it is obtained the well-known expression for the time flight in the ideal case.   \medskip

Another way to analyze the equation \eqref{timeflight1} is by rewriting it as
\begin{equation}\label{timeflight3}
gT_f = \dfrac{kV + g}{k}\left(  1 - e^{-kT_f}       \right) \ \  \  \ \equiv \ \ \ \ h(T_f) = f(T_f),
\end{equation}
where the units of $h$ and $f$ are m/s.
The left-hand side of the last equation ($h(T_f)$) is a linear function whereas the right-hand side ($f(T_f)$) is a transcendental expression. Therefore, it is not possible to get an analytic expression for $T_f$ by means of an easy and simple mathematical procedure. One approach to obtain an approximate solution to  the last equation is using numerical methods.  On the other hand, as we mentioned in the introduction, it is possible to find analytical solutions for the time of flight  in terms of the Lambert $W$ function which is available in some computational tools  as \textit{Geogebra}, \textit{Maple} and \textit{Mathematica}. \medskip

\begin{figure}[h]
\centering
\includegraphics[width=0.5\linewidth]{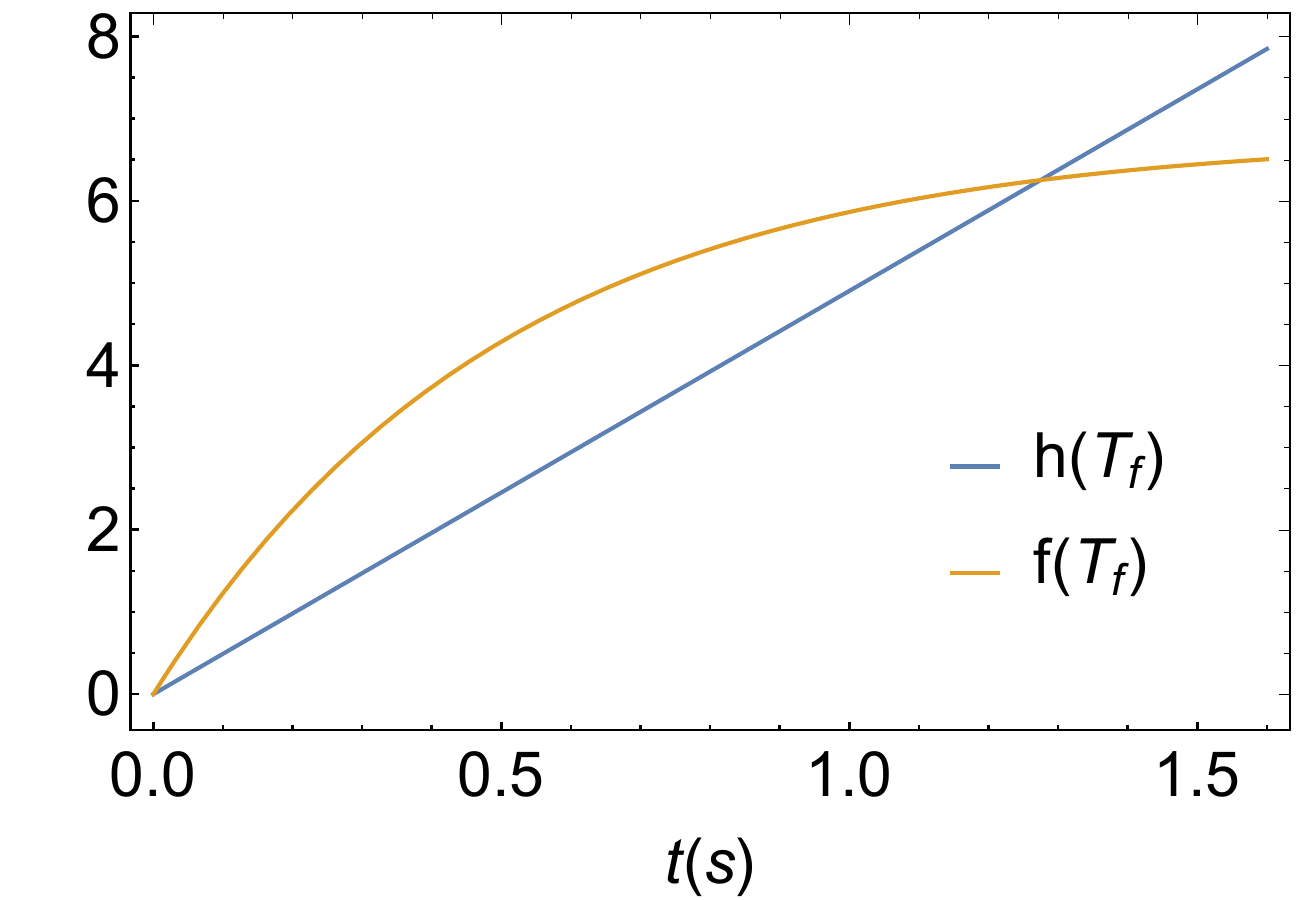}
\caption{The linear $h(t)$ (blue) and the transcendental $f(t)$ (orange) expressions as a function of $t$, given by equation \eqref{timeflight3}, with $k=2$  s$^{-1}$, $V_0 = 10$ m/s and $\theta_0=60^{\circ}$.}
\label{tiros}
\end{figure}

Considering that the procedure to obtain an analytical expression for $T_f$ in terms of the Lambert $W$ function is difficult,  abstract  and cumbersome for undergraduate students from different fields other than physics or mathematics, or those who have not taken a course on special functions, we use the  Plot command in \textit{Mathematica}, without employing the Lambert $W$ function, to obtain the flight time easily. We get numerical values for $T_f$  taking the coordinates of the intersection point, namely,  where $h(T_f) = f(T_f)$,  for several values of the retarding force parameter $k$ assuming that   the launch angle is fixed.  Figure \ref{tiros}  shows the graphs of $h(T_f)$ and $f(T_f)$ (see  equation \eqref{timeflight3}) with $k = 2$  s$^{-1}$ and $\theta_0=60^{\circ}$. In this case, we obtain that the solution is given by $T_f = 1.274$  s.   \medskip

\begin{figure}[H]
\centering
\includegraphics[width=0.5\linewidth]{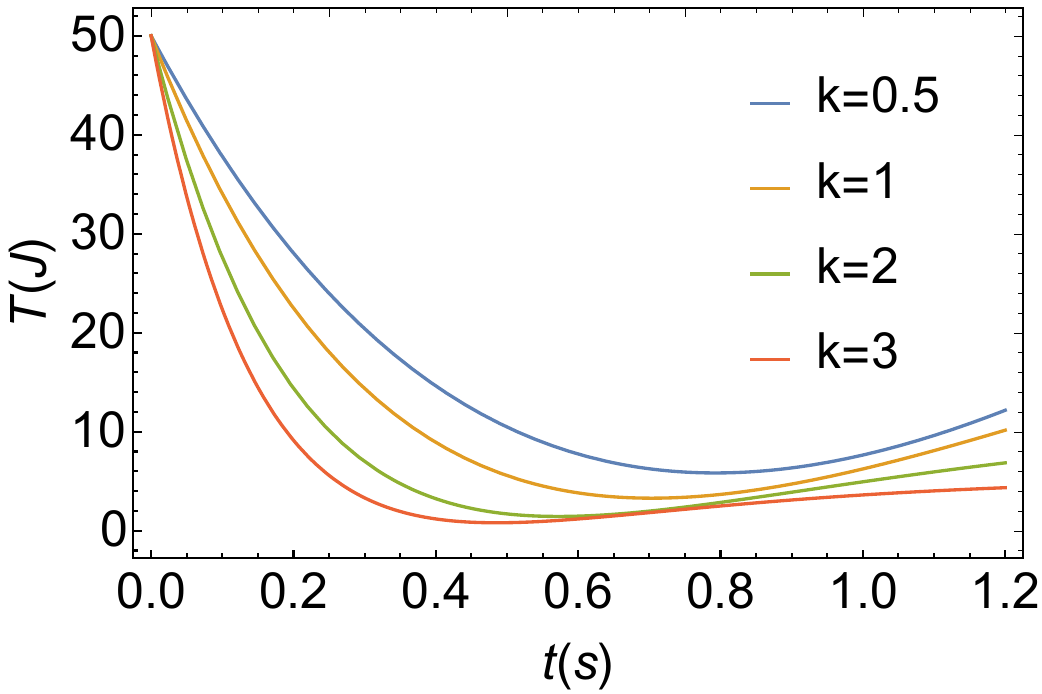}
\caption{Kinetic energy of the projectile with $V_0 = 10$ m/s,  $\theta_0=60^{\circ}$ and different values of the parameter $k$ in  s$^{-1}$. }
\label{energies}
\end{figure}

Now, we are going to analyze the effect of the parameter $k$ on  kinetic, potential and total energies.  Figure \ref{energies}  shows the kinetic energy $T = \frac{1}{2}  m (\dot{x}^{2} + \dot{y}^{2})$  in function of time taking $m = 1$ kg. Similar to the case without friction, kinetic energy has a minimum value, but it is not obtained at the point of maximum height. To obtain the time at which the kinetic energy is minimum, we set $dT/dt = 0$ and obtain
\begin{equation}\label{t-min} 
t_{min} = \dfrac{1}{k} \text{ln} \left[  \dfrac{kV + g}{g}  \left(  1 + \left( \dfrac{kU}{kV +g} \right)^2 \right)   \right]   =  t_{peak} +  \dfrac{1}{k} \text{ ln} \left[  1 + \left( \dfrac{kU}{kV +g} \right)^2 \right].
\end{equation}
Thus, it is demonstrated that the time at which the kinetic energy is minimum $(t_{min})$ is different from the time required to reach maximum height $(t_{peak})$. As $t_{min} \geq t_{peak}$, the kinetic energy is minimum when the projectile is descending.
When the time is bigger than the time for reaching the maximum height $(t \gg   t_{peak})$, the kinetic energy becomes constant because $\dot{x} \rightarrow 0$ (see equation \eqref{velocity-x})
and $\dot{y}  \rightarrow -g/k$ (see equation \eqref{velocity-y}), where $-g/k$ is the  terminal velocity.\medskip

\begin{figure}[H]
\centering
\includegraphics[width=0.5\linewidth]{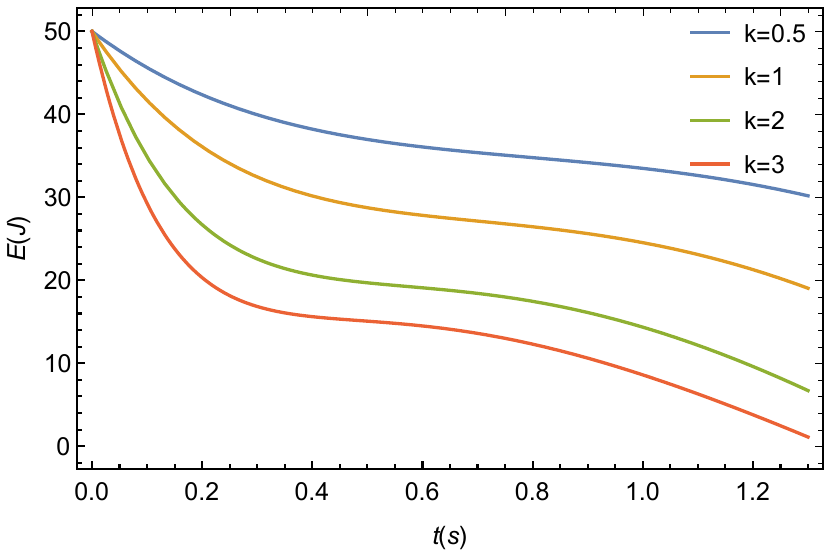}
\caption{Total energy of the projectile with $V_0 = 10$ m/s,  $ \theta_0=60^{\circ}$ and several values of the parameter $k$ in s$^{-1}$. }
\label{Etotal}
\end{figure}

The potential energy $U = m g y$ as a function of time has the same behavior as $y = y(t)$. The only difference between the $U$ vs $t$ and $y$ vs $t$ graphs is the  vertical scale.  Figure \ref{Etotal} shows the total energy $E = T + U$ of the particle in function of time.  At $t=0$, the total energy is $E(t=0) =mV_0^2/2$, so all the curves start from the same point for different values of $k$. In the ideal case, a horizontal line is obtained at this value because during the ascent (descent), potential energy $U$ increases   (decreases ) at the same rate that kinetic energy $T$ decreases (increases).  When there is a resistance medium, the loss of kinetic energy $T$ during ascent is not compensated by an increase in potential energy $U$, resulting in energy dissipation. This dissipation is greater during ascent because it is proportional to the projectile's velocity, causing the curve to have a steeper slope at each point. During descent, the projectile falls more slowly, resulting in less energy loss and a less steep slope of the curve. If the projectile continues its motion below $y=0$, the curve asymptotically tends towards a line with a slope of $-mg^2/k$ (proportional to the terminal velocity). \medskip

The colored curves in figure 3 do not intersect each other for different values of k.  In the case of figure 4,  the colored curves can cross each other for different values of $k$ only when $E <  0$, that is,  for times greater than the time needed to reach the range $R$.\medskip

 For completeness, we consider the rate of  energy-loss $(dE/dt)$  that it is the slope of Figure \ref{Etotal}. It is given by
 \[ \frac{dE}{dt}=  \frac{dT}{dt} + \frac{dU}{dt}= m (\dot{x}\ddot{x} +\dot{y}\ddot{y} ) + mg\dot{y}. \]
For large times $\dot{x} \rightarrow 0$,  $\dot{y} \rightarrow -g/k$, $\ddot{x} \rightarrow 0$  and $\ddot{y} \rightarrow 0$ (see equations \eqref{velocity-x}, \eqref{velocity-y}, \eqref{acceleration} and \eqref{acceleration1}). Thus,  $dE/dt$ is dominated by $dU/dt$ and tends to the constant value of $ -mg^{2}/k$.

Finally, we are going to obtain, in an approximate way, the expressions for $t_{peak}$, $Y_m$, $X_m$ and the trajectory. Using the expansions $   
 \text{ln}  (1+z) = z - z^{2}/2 + z^{3}/3 - ...$  in equations  \eqref{t-subida}, \eqref{Ymax} and \eqref{Eq-trajectory},  and $(1 + z)^{-1} = 1 - z + z^{2} - ...$ in equation \eqref{Xmax}, it is obtained

\begin{eqnarray}
t_{peak} &=& \dfrac{  V}{g} \left[ 1 - \frac{1}{2} \left( \dfrac{kV}{g} \right) + \frac{1}{3}\left( \dfrac{kV}{g} \right)^{2} - \cdots   \right] \\
Y_{m} &=& \dfrac{V^2}{2g}\left[ 1 - \dfrac{2}{3} \left( \dfrac{kV}{g} \right) + \cdots \right]\\
X_{m} &=& \dfrac{ 2UV}{2g}\left[ 1 -  \dfrac{kV}{g} + \left( \dfrac{kV}{g} \right)^{2} - \cdots \right] \\
y &=& (tan  \theta_0)x  - \dfrac{g}{2U^2}x^{2} - \dfrac{gk}{
 3U^3}x^{3} +\cdots,
\end{eqnarray}

where we have assumed that the dimensionless perturbative parameter $\frac{k V}{g}$ is small.  Taking $k = 0$ in the above equations, it is obtained the well-known expressions for the projectile motion in absence of medium resistance.\medskip

We provide \cite{camor}
a didactic simulation in \textit{GeoGebra} to obtain the graphs of all the equations obtained in this section. With this computational tool, the reader can easily verify the presented results in this section by manipulating the values of the parameter $k$, the initial velocity $V_0$, and the launch angle $\theta_0$.

\section{Behavior of the radial distance $(r)$ and the critical angle $(\theta_{c})$}
The radial distance from the origin of the coordinate system to the projectile can be expressed as a function of time ($r = \sqrt{x^{2}(t) + y^{2}(t)}$) or the horizontal distance ($r = \sqrt{x^{2} + y^{2}(x)}$).  Some years ago, Walker \cite{Walker1995} reported an interesting phenomenon in the projectile motion in  absence of a friction force. He found that  projectiles are  "coming and going" for launch angles greater than $70.5288^{\circ}$. From this critical angle ($\theta_{c}$), the radial distance exhibits an oscillation: it increases, decreases and increase again.  Recently, Ribeiro and Sousa  \cite{Ribeiro2021} extended the Walker's work and demonstrated that the "coming and going" phenomenon is also present in the projectile motion with a linear resistance force.  Motivated by the previous works of Walker \cite{Walker1995} and Ribeiro-Sousa \cite{Ribeiro2021} we have scrutinized this fascinating result about the oscillation of the radial distance. \medskip

\begin{figure*}[!t]
	\centering
 \includegraphics[scale=0.4]{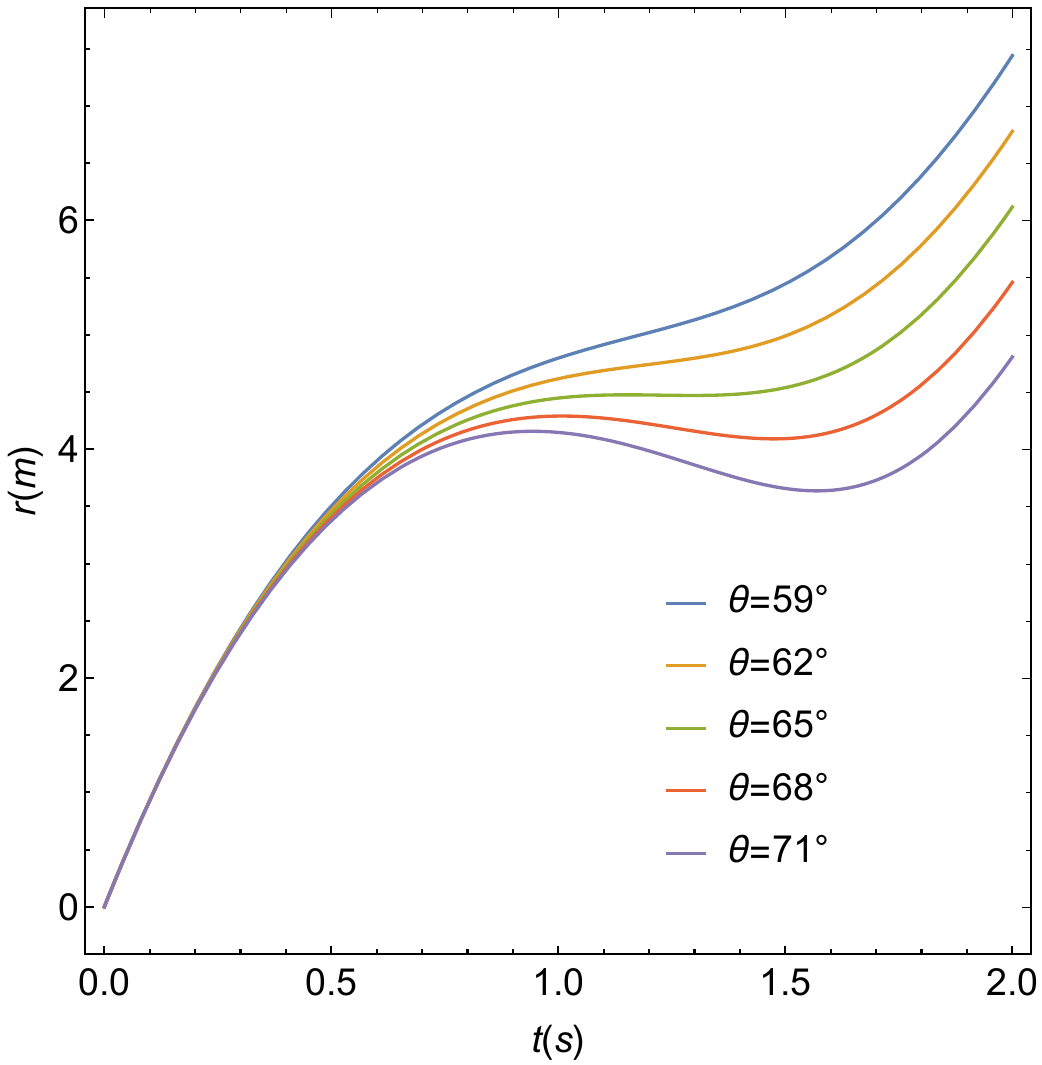} 
  \includegraphics[scale=0.389]{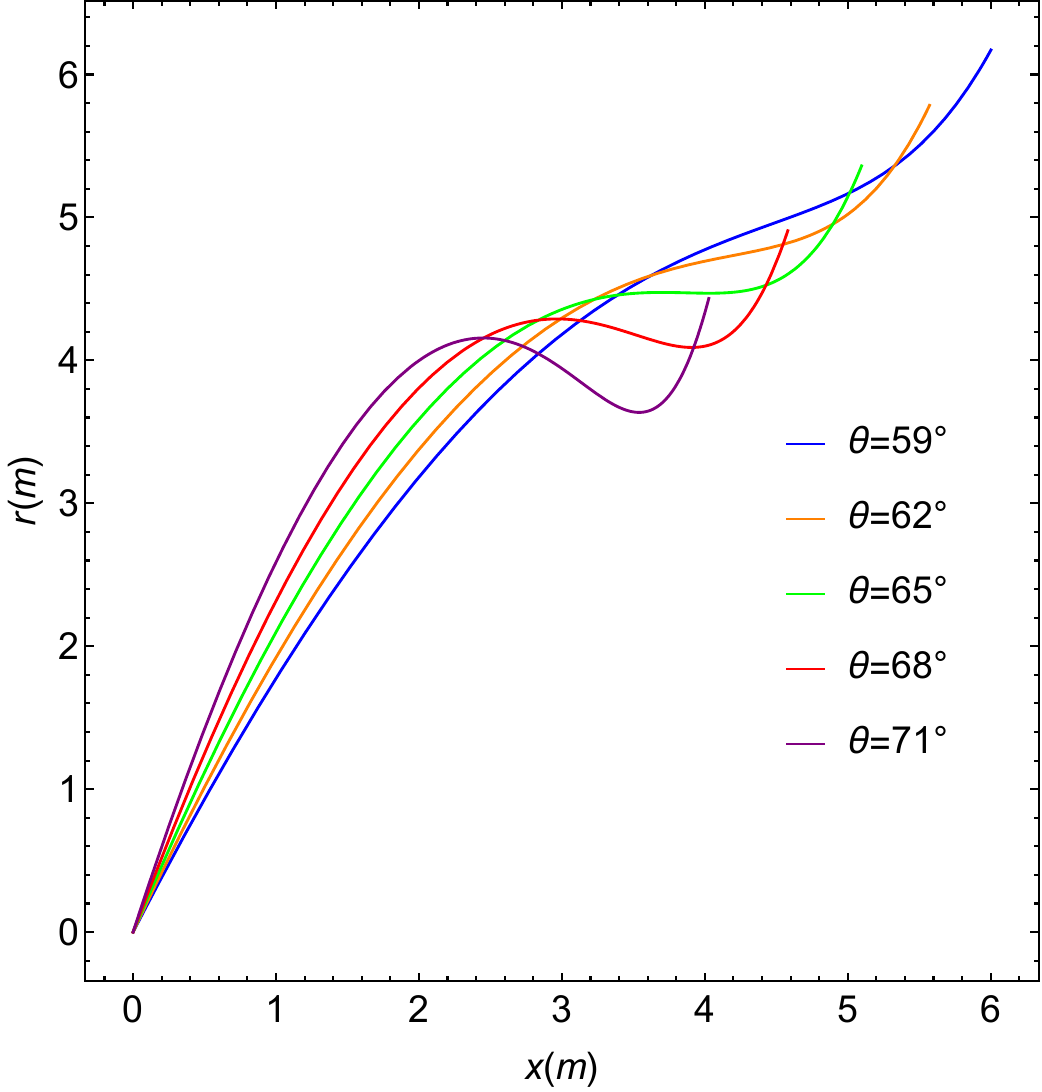}\ 
	\caption{\small Radial distance as a function of time (\textit{left}) and radial distance as a function of horizontal position (\textit{right}) with $k = 0.5$ s$^{-1}$, $V_0 = 10$ m/s and $g= 9.81$  m/s$^2$.}
	\label{radial distance}
\end{figure*}

Figure \ref{radial distance}  shows  $r$ as a function of  $t$ \textit{(left)} and of  $x$ \textit{(right)}, respectively, for several values of the launch angle with $k=0.5$ s$^{-1}$, $V_0 = 10$  m/s and $g = 9.81$  m/s$^2$. In both graphs, a radial oscillation is observed starting from a launch angle close to $65^{\circ}$, approximately.   The radial distance increases, decreases and increases again from this launch angle. \medskip

It is interesting to highlight that the behavior of the radial distance $r$ with respect to the time $t$ and the horizontal distance $x$ is different (see figure \ref{radial distance}). This behavior  depends on the condition $dr/dt$ = 0 and $dr/dx = 0$. They are related by 
$\frac{dr}{dt}= \frac{dr}{dx}\frac{dx}{dt} = \dot{x} \frac{dr}{dx}. $  
When friction is present, $\dot{x}$ depends on time; therefore, the behavior of $dr/dt$ and $dr/dx$ is different. In the opposite case, when there is no friction, $\dot{x}$ is constant. Hence, the behavior of $dr/dt$ and $dr/dx$ is the same. \medskip

Now we are going to obtain, in an approximate way, an expression for the launch critical angle, $\theta_{c}$, from which the radial oscillation is manifested. We begin by requiring that the radial velocity be zero:

\begin{equation}
    \frac{dr}{dt} =  \frac{x \dot{x} + y \dot{y}}{\sqrt{x^{2} + y^{2}}}=0.
\end{equation}

Assuming that $kt \leq 1$, we use the expansion  $e^{-kt} \approx 1 - kt + \dfrac{(kt)^{2}}{2}- \dfrac{(kt)^{3}}{6}$ and  substitute it in the expression $x \dot{x} + y \dot{y}=0$. Thus,  we obtain the condition

\begin{equation}
    (7 k^{2} V_0^{2} + 10 k V g  + 3 g^{2})t^{2}- 9 (k V_0^{2} + Vg) t + 6 V_0^{2} =0.
\end{equation}
Solving for $t$ yields 
\begin{equation}
    t = \frac{9 (k V_0^{2} + Vg) \pm \sqrt{81 (k V_0^{2} + Vg)^{2} - 24 V_0^{2}(7 k^{2} V_0^{2} + 10 k V g  + 3 g^{2})} }{2(7 k^{2} V_0^{2} + 10 k V g  + 3 g^{2})}.
\end{equation}

As the phenomenon of radial oscillation begins when the last equation  has a single solution, it is required that the quantity inside the square root be zero. This gives:

\begin{equation}
27g^{2}\sin^{2}{\theta_c} - 26 k g V_0\sin{\theta_c} - (29 k^{2}V_0^{2} + 24g^{2})=0,
\end{equation}

where we have used $V = V_0 \sin \theta_c$.  Solving the last equation for $\sin \theta_c$, it is obtained

\begin{equation}
\sin \theta_c = \frac{13}{27} \Gamma \pm \frac{2}{9}\sqrt{ \frac{238}{9}\Gamma^{2} + 18 },
\end{equation}

where the dimensionless parameter $\Gamma$ is defined by  $\Gamma \equiv \frac{k V_0}{g}$. The solution with sign minus gives negative angles. So, we take the solution with sign plus and obtain  the expression for the critical angle in function of the parameter $\Gamma$:
\begin{equation}
\theta_c(\Gamma) = \sin^{-1}\left(\frac{13}{27} \Gamma + \frac{2}{9}\sqrt{ \frac{238}{9}\Gamma^{2} + 18 }\right).
\end{equation}

\begin{figure}[h]
\centering
\includegraphics[width=0.6\linewidth]{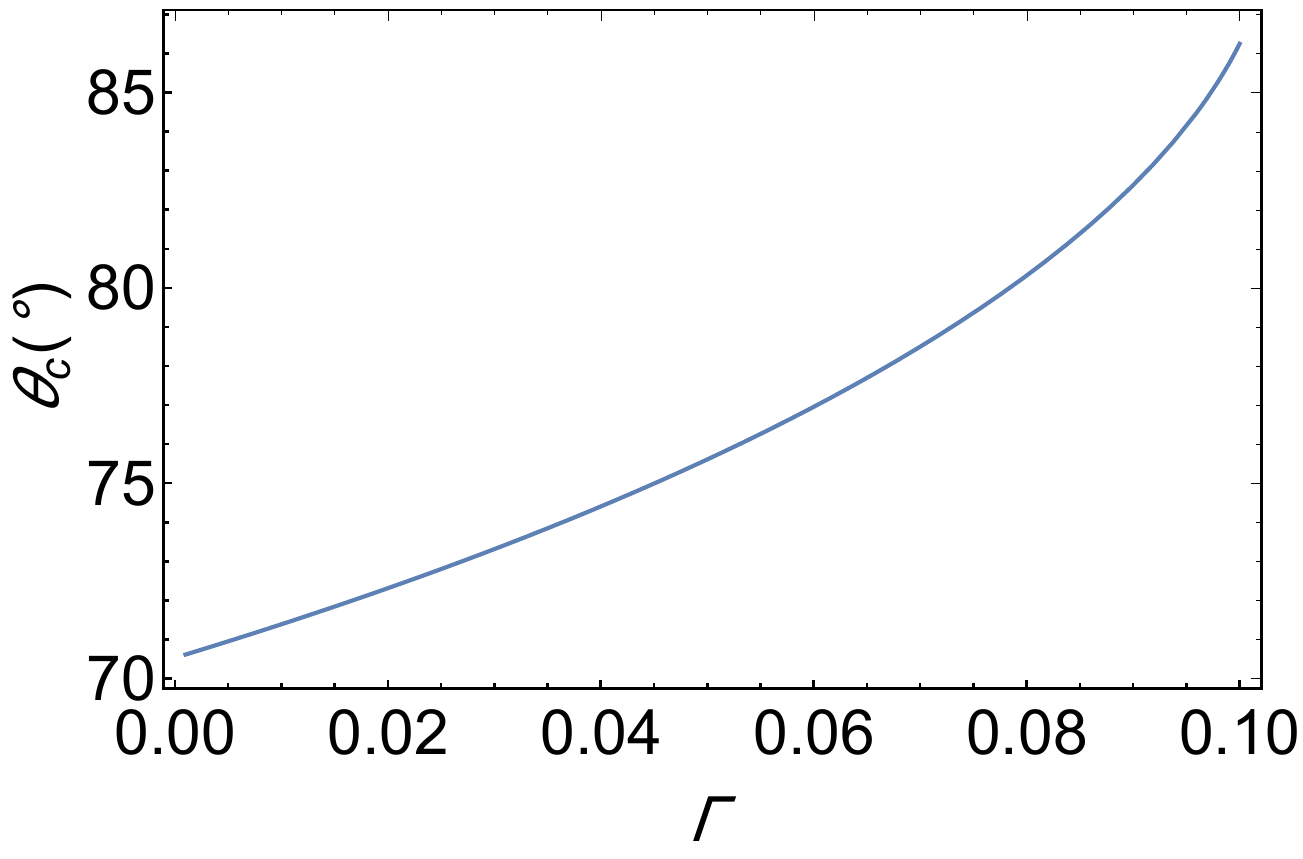}
\caption{Behavior of the  critical angle  in function of $\Gamma$.}
\label{critical angle}
\end{figure}

 This expression is only valid for small values of $k$. Figure \ref{critical angle}  displays the critical angle for small values of the parameter $\Gamma$.
Let us mention that for $k=0$ it is obtained $\theta_c \approx 70.53$. So, this result agrees with the value obtained in Ref. \cite{Walker1995} without the linear resistance force. However, our result is different from the one reported  in the equation (9) of \cite{Ribeiro2021}. According to  this reference, it is reported that $\theta_c(\Gamma) = \sin^{-1}\left(-\frac{5}{27} \Gamma + \frac{2}{9}\sqrt{ \frac{40}{9}\Gamma^{2} + 18 }\right).$   We consider that there is a mistake in this expression which likely  arises from the algebraic manipulation of the Taylor series used to expand $x \dot{x} + y \dot{y}=0$.  \medskip

Finally, let us mention that if we start by demanding that $dr/dx = 0$, where $r = \sqrt{x^2 + y^2(x)}$, we obtain, through a  similar algebraic procedure,  the following expression for the critical angle assuming that $\Gamma \equiv \frac{k V_0}{g}\ll 1$:
\begin{equation}
\alpha_c(\Gamma) = \sin^{-1}\left(-\frac{32}{27} \Gamma + \frac{2}{9}\sqrt{ \frac{256}{9}\Gamma^{2} + 18 }\right).
\end{equation}

\section{Evolution of the apexes}
When the fluid resistance is not considered $(k=0)$, the locus conformed by the set of  points corresponding to the maximum height, $(Y_{m}(\theta_0), X_m(\theta_0))$,  where

\begin{center}$Y_{m}(\theta_0)= (V_0\sin{\theta_0})^{2}/2g$  \ \ \ and \ \ \ $X_m(\theta_0)=(V_0^{2}\sin{2\theta_0})/g $, 
\end{center}
describes an ellipse in function of the launch angle (see, for example,  Refs. \cite{Fowles2005, Walker1995, Chapou2004,  Thomas2004, Soares2013, Rizcallah2020}). If the retarding force is considered, i.e., $k \neq  0$, it is natural to extend the same analysis to this type of motion and pose the following  question:  what is the locus that describes the maxima  of all  projectile trajectories as a function of the launch angle? \medskip

Stewart \cite{Stewart2006} and Hernandez \cite{Hernandez2010} demonstrated by means of the Lambert $W$ function in Cartesian form and polar coordinates, respectively, that the locus of the set of maxima of the  projectile motion  in a linear resisting  medium is not an ellipse. The procedure to  obtain an explicit analytical expression $y = y(x)$ for the locus of the apexes  is not  easy or  trivial  in terms of the Lambert $W$ function. It is rather something tedious and cumbersome. For that reason, we have revisited it and obtained this locus in Cartesian coordinates,  in an easier way,  without using the Lambert $W$ function.  \medskip

In order to derive an expression for the locus of the set of  maxima in Cartesian coordinates without using the Lambert function,  it is necessary to obtain   $Y_{m}$ as a function of $X_{m}$   from equations \eqref{Ymax} and \eqref{Xmax}. To accomplish this,  we substitute $V = V_{0} \sin \theta_0$ into equation \eqref{Xmax} and then clear $ \sin  \theta_0$ in function of $X_{m}$ obtaining, after some algebraic manipulations, the following expression

\begin{equation}
\sin^4{\theta_0} +  \left( \frac{X_m^2 k^2}{V_0^2} - 1   \right)  \sin^2{\theta_0} + \frac{2gkX_m^2}{V_0^3} \sin{\theta_0} + \frac{X_m^2 g^2}{V_0^4}  = 0.
\end{equation}

This is a equation of fourth degree in $\sin{\theta_0}$. With the help of \textit{Mathematica} and some manipulation by hand, we find that the two physical solutions of this  equation, in function of $X_m$, are given by 

\begin{equation}
\sin{\theta_1} = \frac{1}{2}\left(\sqrt{E}-\sqrt{- 6A - E - \frac{4gkX_m^2}{V_0^3 \sqrt{E}}  }\right)
\end{equation}

\begin{equation}
\sin{\theta_2} = \frac{1}{2}\left(\sqrt{E} + \sqrt{- 6A - E - \frac{4gkX_m^2}{V_0^3 \sqrt{E}}  }\right),
\end{equation}
where

\begin{equation}
    A = \frac{k^2 X_m^2 - V_0^2}{3V_0^2}, 
\end{equation}

\begin{equation}
   E =  -2A + \frac{B}{3\left(  C + \sqrt{D + C^2}  \right)^{1/3}}  + \frac{1}{3 (2^{1/3}) V_0^4} \left(  C +  \sqrt{D + C^2 }    \right)^{1/3},
\end{equation}
with 
\begin{equation}
    B = 2^{1/3} \left( V_0^4 - 2 k^2 X_m^2 V_0^2  + 12 g^2 X_m^2 + k^4 X_m^4  \right),  
\end{equation}

\begin{equation}
    C = 108 g^2 k^2 X_m^4 V_0^6 - 72 g^2 X_m^2 V_0^6 \left(  k^2  X_m^2 - V_0^2 \right) + 2V_0^6 \left(k^2  X_m^2  - V_0^2 \right)^3,  
\end{equation}

\begin{equation}
    D = -4 \left( 12 g^2 X_m^2 V_0^4 + V_0^4 (k^2  X_m^2 - V_0^2)^2  \right)^3.  
\end{equation}

After,  we replace the expressions for $\sin {\theta_1}$ and ${\sin\theta_2}$ in equation \eqref{Ymax}. Thus, we obtain in Cartesian coordinates the trajectory of the set of  maxima  of the projectile motion, $Y_{m} (X_{m})$: 

\begin{equation}
Y_{m}(X_m) =  -\dfrac{g}{k^{2}}  \text{ln} \left[ 1 +  \dfrac{kV_{0} \sin \theta_{1(2)}}{g} \right] + \dfrac{V_{0} \sin \theta_{1(2)}}{k}.  
\end{equation}

\begin{figure}[h]
\centering
\includegraphics[width=0.6\linewidth]{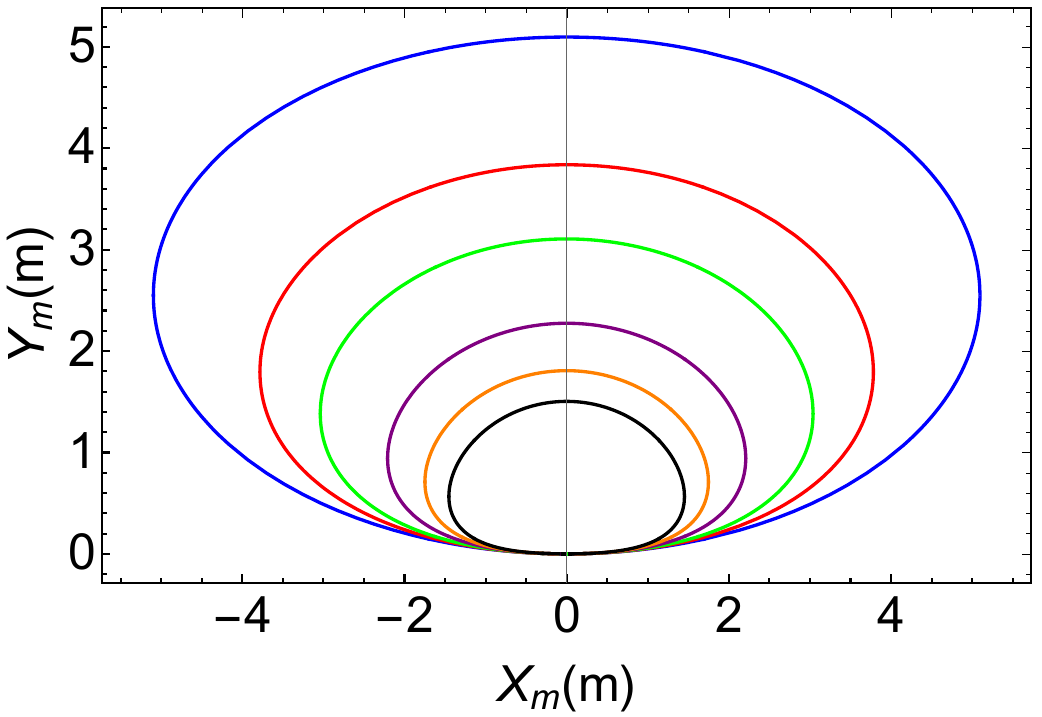}
\caption{Trajectory of the highest point of the projectile for  $k = 0.0001$ s$^{-1}$ (blue), $k = 0.5$ s$^{-1}$ (red), $k = 1$ s$^{-1}$ (green), $k = 2$ s$^{-1}$ (purple), $k = 3$ s$^{-1}$ (orange), and $k = 4$ s$^{-1}$ (black). } 
\label{elipse}
\end{figure}

Figure \ref{elipse}  displays this trajectory for different values of the parameter  $k$ with $V_0 = 10$ m/s. It shows that the trajectory of the peak is not an ellipse when the retarding force $\vec{F_r} = -mk\vec{V}$ is included. The upper part of each graph is obtained replacing $\sin {\theta_1}$ in  $Y_{m}$, while   the lower part of each graph is obtained using $\sin {\theta_2}$ in  $Y_{m}$. We highlight that  these trajectories are obtained without the Lambert $W$ function. \medskip

\section{Summary}
In this article we  presented a didactical overview on the projectile motion considering  a retarding  force proportional to the velocity, $\vec{F_r} = -mk\vec{V}$, where the parameter $k$ gives the strength of  the force. We obtained an analytical expression for the set of maxima of the trajectories, in Cartesian coordinates, without using the Lambert $W$ function, showing that the trajectory of the locus of the apexes in function of the launch angle seems a deformed ellipse.   This calculation complements previous results obtained   in Refs. \cite{Stewart2006, Hernandez2010}, in Cartesian and polar coordinates, but employing the Lambert $W$ function.  According to our literature review, this result has not yet been published.  \medskip

Also, we extended the analysis  performed in  Ref. \cite{Walker1995} for the ideal case $(k=0)$ and complemented the results of Ref. \cite{Ribeiro2021} related with the effect of the parameter $k$ on the radial distance, showing that there is a radial oscillation from certain critical angle which depends of $k$  and obtained, in an approximate way,  the expression for this angle.   \medskip

In our analysis, we have included the impact of the parameter $k$ in the kinetic energy, the potential energy, the total energy and the rate of energy loss, and in the $\dot{x}-x$, $\dot{y}-y$ and $v - r$ phase plots. Although this is straightforward with the help of the \textit{Mathematica} package,  up to now the analysis of these important observables have not been included in the literature. We provide a simulation \cite{camor}, very easy to manipulate, in which the reader can verify all the results and graphs obtained in this work by changing the numerical values of the parameter $k$, the initial velocity $V_0$, and the launch angle $\theta$. \medskip

In general, our results extend the knowledge of the projectile motion assuming a retarding  force proportional to the velocity and can be included in an  intermediate-level  classical mechanics course.  \medskip

\textbf{Acknowledgment}. \\
The authors would like to express their gratitude to the referees for their invaluable comments, which significantly contributed to improving the manuscript.


\begin{thebibliography}{10}

\bibitem{Serway2004} Serway R A and Jewet J W 2004 \textit{Physics for Scientists and Engineers} 6th edn (Belmont: Thomson Brooks/Cole)

\bibitem{Halliday2010} 
Halliday D,  Resnick R and Walker J 2010 \textit{Fundamentals of physics} (New Jersey: Jhon Wiley and Sons)

\bibitem{Tipler2008} Tipler P A  and  Mosca G 2008    \textit{Physics for Scientists and Engineers} 6th edn (New York: W. H. Freeman)
 
\bibitem{Sears2008} Young H D and Freedman R A 2008 \textit{Sear’s and Zemansky’s University Physics: With Modern Physics} 12th edn (Reading, MA: Addison-Wesley) 

 
 


\bibitem{Appell}
Appell P 1941 \textit{Traité de Mécanique Rationnelle}6th edn  (Sceaux: Éditions Jacques Gabay)

\bibitem{Marion-Thornton} Thornton S T  and  Marion J B 2003 \textit{Classical dynamics of particles and systems} 5th edn (Belmont: Thomson Brooks/Cole).

\bibitem{Symon}Symon K R 1971 \textit{Mechanics} 3rd edn (Massachusetts: Addison-Wesley Publishing Company) 

\bibitem{Murphy1972} Murphy R V 1972 Maximum range problems in a resisting medium \textit{The Mathematical Gazette} \textbf{56} 10 

\bibitem{Mestre1990} de Mestre N 1990 \textit{The Mathematics of Projectiles in Sport} (Cambridge: Cambridge University Press)

\bibitem{Lange}
  de Lange O L and  Pierrus J 2010 \textit{Solved Problems in Classical
Mechanics} (New York: Oxford University Press)

\bibitem{Richard}
 Price R H and  Romano J D 1998 Aim high and go far—Optimal projectile launch angles greater than 45° \textit{Am. J. Phys.} \textbf{66}, 109


 
\bibitem{Vennard} Vennard J K 1940 \textit{Elementary fluid mechanics} (New York: John Wiley and Sons, Inc).
 
\bibitem{Bachelor} Batchelor G K 1967 \textit{An Introduction to Fluid Dynamics} (London: Cambridge University Press).

\bibitem{Long-Weiss}Long L N  and Weiss H 1999 The velocity dependence of aerodynamic drag: a primer for mathematicians Am. Math. Monthly, \textbf{106}(2), 127  


\bibitem{Greiner2003} Greiner W 2003 \textit{Klassische Mechanik I} (Frankfurt am Main: Verlag Harri Deutsch)



\bibitem{Erlichson1983} Erlichson H 1983 Maximum projectile range with drag and lift, with particular application to golf \textit{Am. J. Phys}. \textbf{51}(4), 357

\bibitem{Martin1991} Martin P and Puerta J 1991 Two-point fractional approximants  for the motion of a projectile in a resisting medium \textit{Eur. J. Phys}. \textbf{12} 86.

\bibitem{Groetsch1996} Groetsch C W 1996 Tartaglia's inverse problem in a resistive medium \textit{Am. Math. Monthly} \textbf{103}(7) 546 

\bibitem{Groetsch1997} Groetsch C W and Cipra B 1997 Halley's comment - Projectiles with linear resistance \textit{Mathematics Magazine} \textbf{70}(4) 273

\bibitem{Alwis2000}de Alwis T 2000 Projectile motion with Mathematica \textit{Int. J. Math. Educ. Sci. Technol}. \textbf{31} 749

\bibitem{Bruno2002} Bruno A D S and Matos J M O 2002 The projectile path lenght (in Portuguese) \textit{Rev. Bras. Ens. Fis}. \textbf{24}(1) 30 



\bibitem{Groetsch2003}Groetsch C W 2003 Timing is everything: The French connection \textit{Am. Math. Monthly} \textbf{110}  950

\bibitem{Groetsch2005}Groetsch C W 2005 Another broken symmetry \textit{Coll. Math. J}. \textbf{36}(2) 109

\bibitem{Fowles2005}  Fowles G R and  Cassiday G L 2005 \textit{Analytical Mechanics}, 7th edn (Belmont: Thomson Brooks/Cole)

\bibitem{Pereira2008}Pereira L R and Bonfim V 2008 Security regions in projectile motion (in Portuguese) \textit{Rev. Bras. Ens. Fis.} \textbf{30}(3) 3313

\bibitem{Borghi2013}Borgui R 2013 Trajectory of a body in a resistant medium: an elementary derivation \textit{Eur. J. Phys}. \textbf{34} 359.


\bibitem{Andersen2015}
 Andersen P W 2015 Comment on ‘Wind-influenced projectile
motion’ \textit{Eur. J. Phys.} \textbf{36}  068003 

\bibitem{Kantrowitz2015}
Kantrowitz R and Neumann M M 2015 Optimization of projectile motion under linear air resistance  \textit{Rend. Circ. Mat. Palermo}  \textbf{64} 365 

\bibitem{Grigore2017} Grigore I, Miron C and Barna E S 2017 Exploring excel spreadsheets to simulate the projectile motion in the gravitational field \textit{Romanian Reports in Physics} 69(1) 

\bibitem{Rizcallah2018}
Rizcallah J A 2018 Approximating the linearly
impeded projectile by a tilted idealized one  \textit{Phys. Educ.} \textbf{53}  065012


\bibitem{Pispinis2019}
Pispinis D 2019 Calculation of minimum speed of projectiles under linear resistance using the geometry of the velocity space \textit{European Journal of Physics Education} \textbf{10} (3)  1-9 

\bibitem{Sarafian2021}
Sarafian, H 2021 What Projective Angle Makes the Arc-Length of the Trajectory in a Resistive Media Maximum? A Reverse Engineering Approach \textit{American Journal of Computational Mathematics} \textbf{11}, 71-82

\bibitem{Ribeiro2021}

Ribeiro W J M and de Sousa J R 2021 Projectile Motion: The ''Coming and Going'' Phenomenon \textit{Phys. Teach.} \textbf{59} 168

\bibitem{Hernandez2022}
Hernandez-Saldaña H 2022 Analytical velocity hodograph of projectile motion under polynomial drag  \textit{Journal of Physics: Conference Series} \textbf{2307}  012019



\bibitem{Warburton2004}Warburton R D H  and Wang J 2004 Analysis of asymptotic projectile motion with air resistance using the Lambert W function \textit{Am. J. Phys}. \textbf{72} 1404.

\bibitem{Packel2004} Packel E W and  Yuen D S 2004  Projectile motion with resistance and the Lambert function \textit{Coll. Math. J}. \textbf{35} 337.

\bibitem{Stewart2005} Stewart S M 2005 Linear resisted projectile motion and the Lambert W function \textit{Am. J.  Phys}. \textbf{73} 199 

\bibitem{Stewart2005A} Stewart S M 2005 A little introductory and intermediate physics with the Lambert W function Proc. 16th Biennial Congress of the Australian Institute of Physics vol 2 ed M Colla (Parkville: Australian Institute of Physics) pp 194–7 

\bibitem{Morales2005}  Morales D A 2005 Exact expressions for the range and the optimal angle of a projectile with linear drag \textit{Can. J. Phys}. \textbf{83} 67 

\bibitem{Stewart2006} Stewart S M 2006 An analytic approach to projectile motion in a linear resisting medium \textit{Int. J.  Math. Educ.  Sci. Technol}. \textit{37}:4 411

\bibitem{Kantrowitz2008} Kantrowitz R and Neumann M M 2008 Optimal angles for launching projectiles: Langrange Vs. CAS \textit{Canad. Appl. Math. Quart.} \textbf{16}(3), 279

\bibitem{Karkantzakos2009} Karkantzakos P A 2009 Time of flight and range of the motion of a projectile in a constant gravitational field under the influence of a retarding force proportional to the velocity \textit{J.  Eng. Sci. Tech. Rev.} \textbf{2} (1)  76.

\bibitem{Hernandez2010} Hernandez-Saldaña H 2010 On the locus formed by the maximum
heights of projectile motion with air resistance \textit{Eur. J. Phys}. \textbf{31} 1319.

\bibitem{Stewart2011Comment} Stewart S M 2011 Comment on 'On the locus formed by the maximum heights of projectile motion with air resistance' \textit{Eur. J. Phys.} 32 L7  

\bibitem{Hernandez2011Reply} Hernandez-Saldaña H 2011 Reply to 'Comment on "On the locus formed by the maximum heights of projectile motion with air resistance"'
 \textit{Eur. J. Phys.} 32 L11 

\bibitem{Stewart2011} Stewart S M 2011 Some remarks on the time of flight and range of a projectile in a linear resisting medium \textit{J.  Eng. Sci.  Technol. Rev}. \textbf{4} (1)  32
 
\bibitem{Morales2011}Morales D A 2011 A generalization on projectile motion with linear resistance \textit{Can. J. Phys}. \textbf{89} 1233

\bibitem{Stewart2012} Stewart S M  2012  On the trajectories of projectiles depicted in early ballistic woodcuts \textit{Eur. J. Phys}. \textbf{33}  149

\bibitem{Hu2012}Hu H, Zhao Y P, Guo Y J and Zheng M Y 2012 Analysis of linear resisted projectile motion using the Lambert W function \textit{Acta Mech}. \textbf{223} 441.

\bibitem{Bernardo2015}Bernardo R C, Esguerra J P, Vallejos J D and Canda J J 2015 Wind-influenced projectile motion \textit{Eur. J. Phys}. \textbf{36} 025016.

\bibitem{Morales2016} Morales D A 2016 Relationships between the optimum parameters of four projectile motions \textit{Acta Mech}. \textbf{227} 1593.


\bibitem{Walker1995}  Walker J S 1995 Projectiles: Are they coming or going? \textit{Phys. Teach.} \textbf{33} 282

\bibitem{camor}
Morales C A, Muñoz J H and C E Vera (2023), \url{https://www.geogebra.org/u/camoralesr}

\bibitem{Chapou2004}  Fern\'andez-Chapou J L,  Salas-Brito A L and  Vargas C A 2004 An elliptic property of parabolic trajectories \textit{Am. J. Phys}. \textbf{72} 1109

\bibitem{Thomas2004} Thomas G B,  Weir M B, Hass J  and  Giordano F R 2004 Calculus 11th edn, p. 930 (Reading, MA: Addison-Wesley)

\bibitem{Soares2013} Soares V,  Tort A C and  de Oliveira Goncalves A G 2013 A note on the parabolic motion: Unexpected circle and ellipse \textit{Rev. Bras. Ens. Fis}. \textbf{35} 2701 
 

\bibitem{Rizcallah2020}
Rizcallah J A 2020 On the elliptic locus of a family of projectiles  \textit{Eur. J. Phys.} \textbf{41}  035004



\end{thebibliography}
\end{document}